\documentclass[3p,twocolumn,sort,compress]{elsarticle}
\usepackage{amsmath}
\usepackage{amssymb}
\usepackage{color}
\newcommand {\GH}[1]{{#1}}

\begin{document}
\begin{frontmatter}
   \title{The influence of substrate temperature on growth of
   para-sexiphenyl thin films on Ir\{111\} supported graphene studied by LEEM}
   \author[UT]{Fawad S. Khokhar}
   \author[UT,MUL]{Gregor Hlawacek\corref{cor3}}
   \ead{g.hlawacek@tnw.utwente.nl}
   \author[UT]{Raoul van Gastel}
   \author[UT]{Harold J. W. Zandvliet}
   \author[MUL]{Christian~Teichert}
   \author[UT]{Bene Poelsema}
   \cortext[cor3]{Corresponding author}

   \address[UT]{University of Twente, MESA$^+$ Institute for Nanotechnology,
   P.O. Box 217, NL-7500AE, Enschede, The Netherlands}
   \address[MUL]{Montanuniversitaet Leoben, Institute for Physics,
   Franz-Josefstrasse 18, A-8700, Leoben, Austria}

   \begin{abstract} 
   
   The growth of para-sexiphenyl (6P) thin films as a function of substrate
   temperature on Ir\{111\} supported graphene flakes has been studied in
   real-time with Low Energy Electron Microscopy (LEEM). Micro Low Energy
   Electron Diffraction ($\mu$LEED) has been used to determine the structure
   of the different 6P features formed on the surface. We observe the
   nucleation and growth of a wetting layer consisting of lying molecules in the initial stages of growth.
   Graphene defects -- wrinkles -- are found to be preferential sites for
   the nucleation of the wetting layer and of the 6P needles that grow on top of
   the wetting layer in the later stages of deposition. The molecular
   structure of the wetting layer and needles is found to be
   similar. As a result, only a limited number of growth directions are observed
   for the needles. In contrast, on the bare Ir\{111\} surface 6P
   molecules assume an
   upright orientation. The formation of ramified islands is observed on the bare Ir\{111\} surface at
   320\,K and 352\,K, whereas at 405\,K the formation of a continuous layer of
   upright standing molecules growing in a step flow like manner is
   observed.

   \end{abstract}


\end{frontmatter}

\section{Introduction}

In recent years, the growth of organic semiconductors on solid substrates
has received significant attention for both scientific and technological
reasons. One such organic semiconductor is para-sexiphenyl (6P), a rigid
rod-like conjugated molecule. Thin film growth of 6P molecules has been
investigated intensely due to the unique optical and electronic properties
of the molecule. These properties are found to be subject to substrate
anisotropy and also depend on the arrangement of the molecules in a thin
film~\cite{Niko1996,Mikami1998}. The molecular orientation can be controlled
by using appropriate substrates from lying~\cite{Teichert2006} to upright
standing~\cite{Hlawacek2008}.
In-depth knowledge of the growth behavior as a
function of temperature is a key to controlling the thin film structure and
exploiting its full technological potential~\cite{Yang2005}. In several recent
publications it has been shown how the growth parameters can be used to
tailor the morphology of 6P thin films on different
substrates~\cite{Kintzel2004,Mullegger2007,Lengyel2008}. In this paper, we
investigate the growth and structure of 6P molecules at different surface
temperatures on epitaxially grown graphene sheets supported by an
Ir\{111\} surface. The layers and needles that form on graphene as well as the
ramified structures that grow on Ir\{111\} are studied as a function of substrate
temperature. The role of defects in the graphene sheets is also
analyzed using Low Energy Electron Microscopy (LEEM) and
Photoemission Electron Microscopy (PEEM). 
Micro Low Energy Electron 
Diffraction ($\mu$LEED) is used to locally obtain
structural information~\cite{Bauer1998}.
 
\section{Experimental}

The experiments are carried out in an Elmitec LEEM III apparatus of Bauer's
design~\cite{Bauer1994} with a base pressure of less than
1$\times$10$^{-10}$\,mbar. A 1.4\,$\mu$m field-limiting aperture has been
utilized to collect local structural information from features of interest.  
An Ir\{111\} substrate is atomically cleaned by
exposing to low pressures of O$_2$ at elevated temperature. Graphene films
are then prepared by Chemical Vapor Deposition (CVD) of ethylene
(C$_2$H$_4$) on the Ir\{111\} surface at a temperature of 875
K~\cite{Coraux2009}. The growth of the graphene flakes is followed in-situ
using PEEM until sufficiently large graphene flakes have formed on the
Ir\{111\} surface. A LEEM image of such a flake is shown in
Fig.~\ref{fig1}(a). Substrate steps (thin lines, indicated by white
arrows) are still visible
in Fig.~\ref{fig1}(a) as the graphene flake follows the topographic contours
of the underlying substrate. A network of straight linear features
(indicated by black arrows), appearing much darker and wider than the
steps, is also visible on the graphene. These linear features are wrinkles
in the graphene sheet that result from elastic relaxations that occur when
the sample is cooled from the graphene growth temperature to the 6P
deposition temperature. The wrinkles extend about 3\,nm from the surface and
are a few nanometers in width~\cite{N'Diaye2009a}. Commercially available 6P
molecules in powder form are deposited by Organic Molecular Beam Epitaxy
(OMBE) from a Knudsen-cell type evaporator that is held at a temperature of
553\,K for all described experiments. From previous experiments it 
was calibrated to yield an
average growth rate of 6.3$\times10^{-4}$\,6P/(nm$^2$s). This corresponds to a growth
rate of 2.7\,ML/h of flat lying 6P molecules~\cite{Hlawacek2011}.
We use the term monolayer for a closed layer of molecules having the
mentioned structure. The number of 6P molecules per surface atom varies
between 0.015\,6P$\mathrm{(1\overline{11})}$/graphene and
0.28\,6P(100)/Ir\{111\}, consequently only deposition times and molecular
densities are given. 
\GH{The sample temperature during deposition of 6P has been varied between
320\,K and 405\,K. In what follows we will refer to this as the deposition
temperature.} These deposition temperatures are
precise relative to each other. However, thermal effects in the sample
holder might lead to a small but unknown offset of all temperatures
given throughout the text.

\section{Results and discussion}

\subsection{Deposition of 6P at 320\,K}
\label{ONE}

\begin{figure} 
   \centering 
   \includegraphics[width=7.5cm]{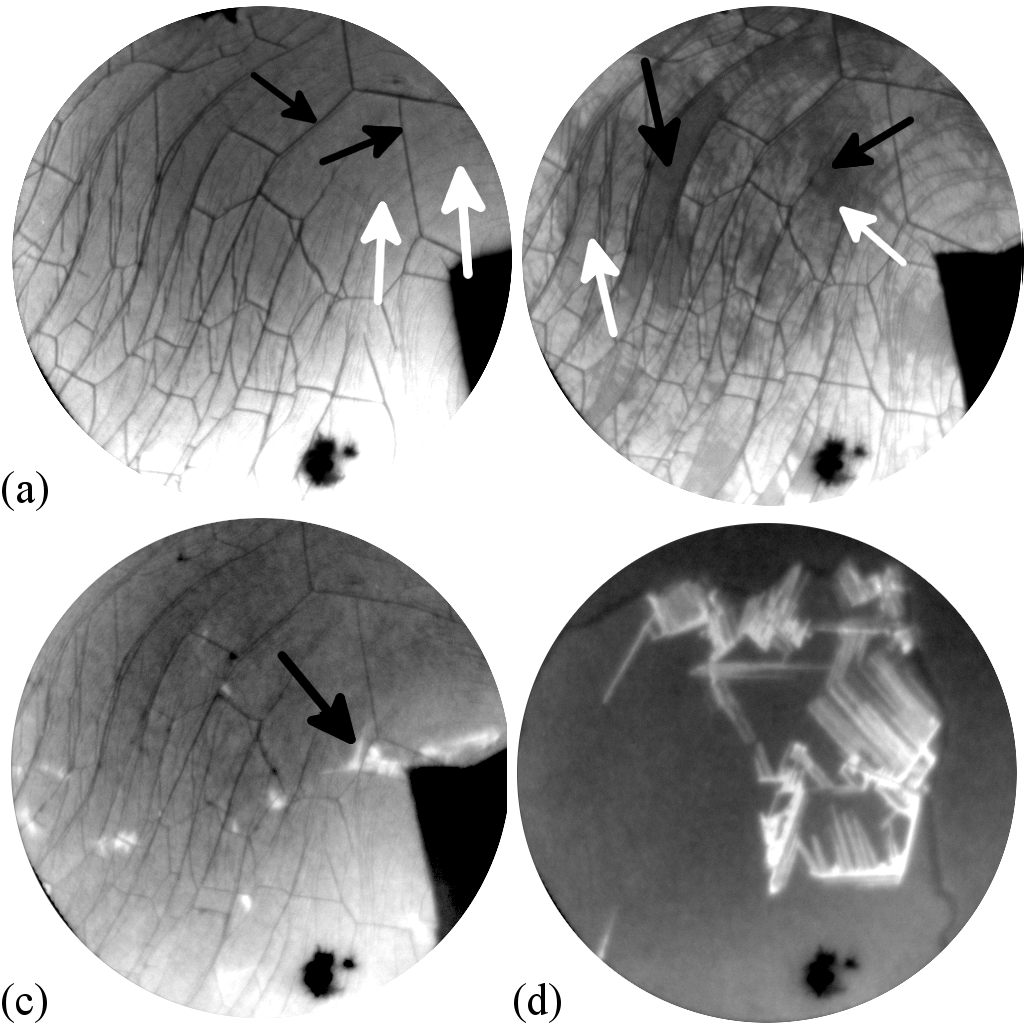} 
   \caption{LEEM images acquired at a
   temperature of 320\,K with an electron energy of 2.7\,eV for (a)-(c)
   and 3.7\,eV for (d). The Field of View (FoV) is 6\,$\mu$m for (a)-(c) and
      4\,$\mu$m for (d). Times indicated are
   measured with respect to the start of 6P deposition. {\bf(a, t=0\,s)} A
   single graphene flake on the Ir\{111\} surface is imaged prior to exposure
   to 6P. Graphene wrinkles (indicated with black arrows) and the faint contours of Ir\{111\}
   surface steps (indicated with white arrows) are visible on the single
   layer graphene flake. {\bf(b, t=813\,s, 0.52\,6P/nm$^2$)} The graphene flake is covered by
   a wetting layer of 6P. The two additional grey levels correspond to the initial
   layer formed by flat lying face-on molecules only (white arrows), and the
   final wetting layer with a \GH{face-on/edge-on}, 
   $(\mathrm{1\overline{11}})$ like structure (black arrows). The
   nucleation of this film happens next to the wrinkles. {\bf(c, t=1268\,s,
   0.80\,6P/nm$^2$)}
   Bright crystallites occur on top of the wetting layer next to the
   wrinkles (black arrow). {\bf(d, t=2149\,s, 1.36\,6P/nm$^2$)}
   Parallel needles continue to grow with ongoing deposition.
   The dark area in
   the lower part of the images is a defect in the channel plate.}
   \label{fig1} 
\end{figure}

A sequence of bright field LEEM images acquired during the deposition of 6P
molecules is shown in Fig.~\ref{fig1}. Figure~\ref{fig1}(a) shows the
pristine graphene surface with graphene wrinkles (thick straight lines) and
steps in the underlying Ir\{111\} surface (thin curved
lines). For a detailed discussion of the morphology of graphene flakes on
Ir\{111\} the reader is referred
to~\cite{N'Diaye2009a,Gastel2009,Coraux2009}. With the deposition of 6P
molecules, the intensity of reflected electrons from the graphene decreases,
indicating the presence of a diluted phase of 6P molecules on the surface.
After 727\,s (0.46\,6P/nm$^2$) of
deposition, nucleation of 6P domains takes place next to the wrinkles. The
domains are mobile and move over the graphene surface~\cite{Hlawacek2011a}.
After 813\,s (0.52\,6P/nm$^2$) of deposition, the intensity that is measured on 6P domains
reduces even further (indicated by black arrows in Fig.\,\ref{fig1}(b)). The
dark 6P domains grow to form a complete monolayer after 948\,s
(0.60\,6P/nm$^2$) of 6P deposition.
For the next 130\,s no new features or significant contrast changes are
observed. After this, bright 6P crystallites can be
observed. These crystallites also nucleate next to the wrinkles,
as indicated by the black arrow in Fig.\,\ref{fig1}(c). In contrast to the
initial islands, these crystals are immobile. With continued deposition,
they elongate, resulting in a fiber like morphology. Fig.\,\ref{fig1}(d) shows a LEEM image after
stopping 6P growth at 2149\,s (1.36\,6P/nm$^2$). The graphene
surface is covered by a 6P wetting layer of monolayer thickness and long fiber-like structures,
which nucleated either from defects in the wetting layer caused by the
wrinkles, or from other needles.

\begin{figure}
   \includegraphics[width=7.6cm]{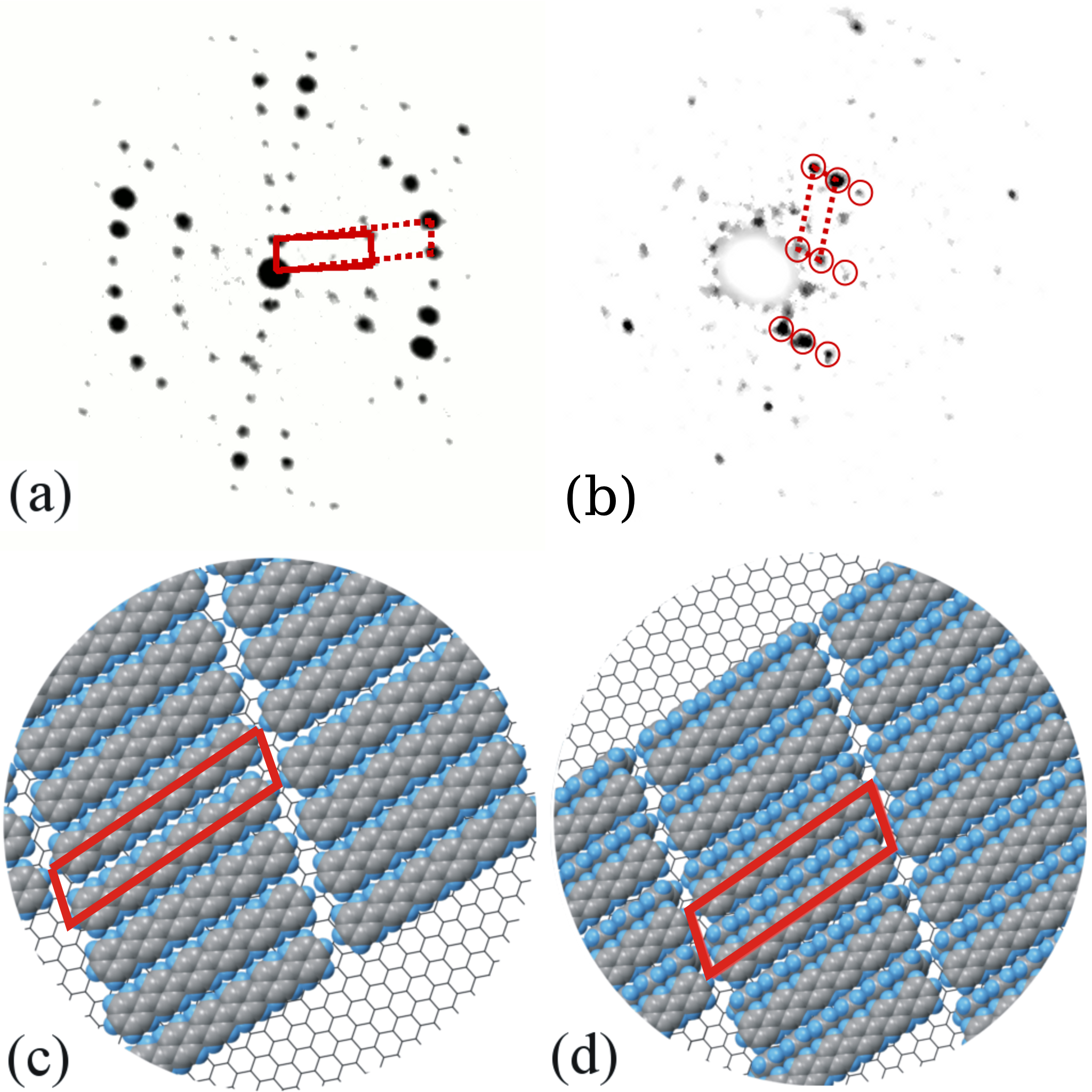}
   \caption{{\bf(a)} $\mu$LEED pattern measured from graphene covered with
   one monolayer of 6P at an electron energy of 14\,eV. The specular
   reflection and other LEED spots associated with various rotational
   domains of the ordered 6P structure are visible. {\bf(b)} $\mu$LEED
   pattern measured from a graphene area covered by needles at an
   electron energy of 21\,eV. The LEED spots are marked with red circles to
   guide the eyes. {\bf(c)} Molecular arrangement
   corresponding to the dashed unit cell in (a). The unit cell contains one
   face-on molecule. {\bf(d)} Sketch of the molecular arrangement corresponding to the
   solid unit cell in (a). Two molecules per unit cell in an alternating
   face-on -- edge-on configuration are found here. The molecular
   arrangement in the needles (b) is similar to this second denser phase present in
   the wetting layer.} 
   \label{fig2}
\end{figure}

Figure~\ref{fig2}(a) shows a $\mu$LEED pattern that is obtained from an area
without needles which is only covered by the monolayer thick wetting layer. The $\mu$LEED pattern consists of the specular
reflection surrounded by several rings of LEED spots. It reveals an ordered
molecular structure. Within the 1.4\,$\mu$m aperture that we used to obtain the
$\mu$LEED pattern several different rotational domains are present. Careful
analysis of the $\mu$LEED pattern also shows that the 6P molecules are
arranged in two different ways, in other words there are two different phases
present. The unit cells are highlighted with solid and dotted lines. The
length of the unit cell vectors, highlighted with dashed lines, are
5.2\,\AA{} 
and 27.8\,\AA{} at an angle $\beta$ of 72$^\circ$. Here, $\beta$ is the angle
between the two lattice vectors. The angle $\Theta$ between the long axis of the 6P
unit cell and the graphene unit cell vector is 79$^\circ$. The dimensions of the unit cell vectors,
highlighted with solid lines, are 8.3\,\AA{} and 27.8\,\AA{} at an angle
$\beta$ of 70$^\circ$. \GH{Taking into account distortions in the LEED pattern
this numbers are accurate within 5\%.} In accordance with the results obtained at 240\,K~\cite{Hlawacek2011} we assume that the first small unit cell
contains one molecule in a face-on configuration (Fig.~\ref{fig2}(c)), while
the second larger unit cell contains two molecules which are assembled in a
face-on -- edge-on arrangement (Fig.~\ref{fig2}(d)). The latter arrangement
is similar to the one found in the surface unit cell of the bulk
6P$(\mathrm{1\overline{11}})$ plane~\cite{bakandfra93}. Also the size of
the unit cell is similar to the bulk surface unit cell. However, the
underlying substrate does not allow the film to relax completely. This results
in a larger spacing along the long molecular axis. \GH{We obtain the
following matrix notations for the unit cell vectors of the adsorbate lattice in
terms of the substrate lattice vectors (a=b=2.46\,\AA{} and
$\alpha$=120$^\circ$):
for the inital layer $\left(\begin{smallmatrix}8.6& 12.8\\-1.3& 1.2  
\end{smallmatrix}\right)$ while the final bulk like layer has a matrix
notation of $\left(\begin{smallmatrix}8.6&12.8\\-1.9&
   2.0\end{smallmatrix}\right)$. These latter values show a good match with
   structural data ($\left(\begin{smallmatrix}
      8.7& 13.0\\-1.7& 1.9 \end{smallmatrix}\right)$ for the final layer)
      obtained at a much lower temperature of 240\,K~\cite{Hlawacek2011}.
      The fact that this
   relationship between the 6P layer and graphene does not change over a
   temperature range of at least 80\,K is a strong hint towards a fixed
   relationship between the two. Keeping in mind the accuracy of our initial
   measurements we therefore interpret this as a
   coincidence type II quasiepitaxial relationship~\cite{Hooks2001}. In fact a 5$\times$10
   superstructure describes the layer more accurately. 
Taking into account the superstructure we arrive at the following matrix notations for the initial $\left(\begin{smallmatrix}43&64\\
-13&12\end{smallmatrix}\right)$
and the final monolayer thick wetting layer
$\left(\begin{smallmatrix}43&64\\
-19&20\end{smallmatrix}\right)$. This also better reflects the fact that in
the superstructure the flexible molecules are free to relax their
orientation and position in the superstructure by small amounts.}

A typical $\mu$LEED pattern taken from needles is shown in
Fig.~\ref{fig2}(b). It consists of LEED spots
from a single domain and thus reveals an ordered molecular
structure. The dimensions of the unit cell vectors are 9.5\,\AA{} and
26.9\,\AA{} at an angle $\beta$ of 69$^\circ$.
The molecular arrangement is
similar to the second denser phase found in the wetting layer
(fig.~\ref{fig2}(d)). Again these values are very
similar to the size of the 6P$(\mathrm{1\overline{11}})$ surface unit cell
and the size of the bigger unit cell found in the wetting layer. However,
the three dimensional shape of the fiber crystallites allows the unit cell
to relax towards the bulk value.

The growth of 6P on graphene at 320\,K can be summarized by the following
four steps. (1) An initial layer of only flat lying molecules is formed on
the graphene surface. This layer nucleates next to the wrinkles. (2) When a
critical coverage is reached, the initial layer transforms into a bulk like
layer (Fig.~\ref{fig1}(b)). The molecules obtain a flat face-on -- edge-on
configuration similar to the 6P$(\mathrm{1\overline{11}})$ plane. (3) 6P
fibers nucleate on top of the monolayer thick wetting layer
(Fig.~\ref{fig1}(c)). This
nucleation occurs next to the wrinkles.
(4) Parallel bundles of needles grow away from the wrinkles
(Fig.~\ref{fig1}(d)). The needles have the same
$(\mathrm{1\overline{11}})$ orientation as the underlying wetting layer.
\GH{The azimuthal orientation of the long needle axis is roughly perpendicular to the
azimuthal orientation of the long unit cell axis and the long molecular axis.}

Nearly all nucleation events are occurring next to the wrinkles. The change
in curvature of the graphene next to the wrinkle, strain in the adsorbed 6P
islands, and the high mobility are responsible for the preferred nucleation
of the wetting layer next to wrinkles and the observed large domain size, which is in
the $\mu$m range. The preferred nucleation, mobility, and formation of the
initial wetting layer of 6P on graphene is discussed in detail
elsewhere~\cite{Hlawacek2011a,Hlawacek2011}. The wrinkles -- by creating a
large network of 1D defects in the 6P wetting layer -- are responsible for
the preferred nucleation of the needles next to them.

\begin{figure}
   \includegraphics[width=7.6cm]{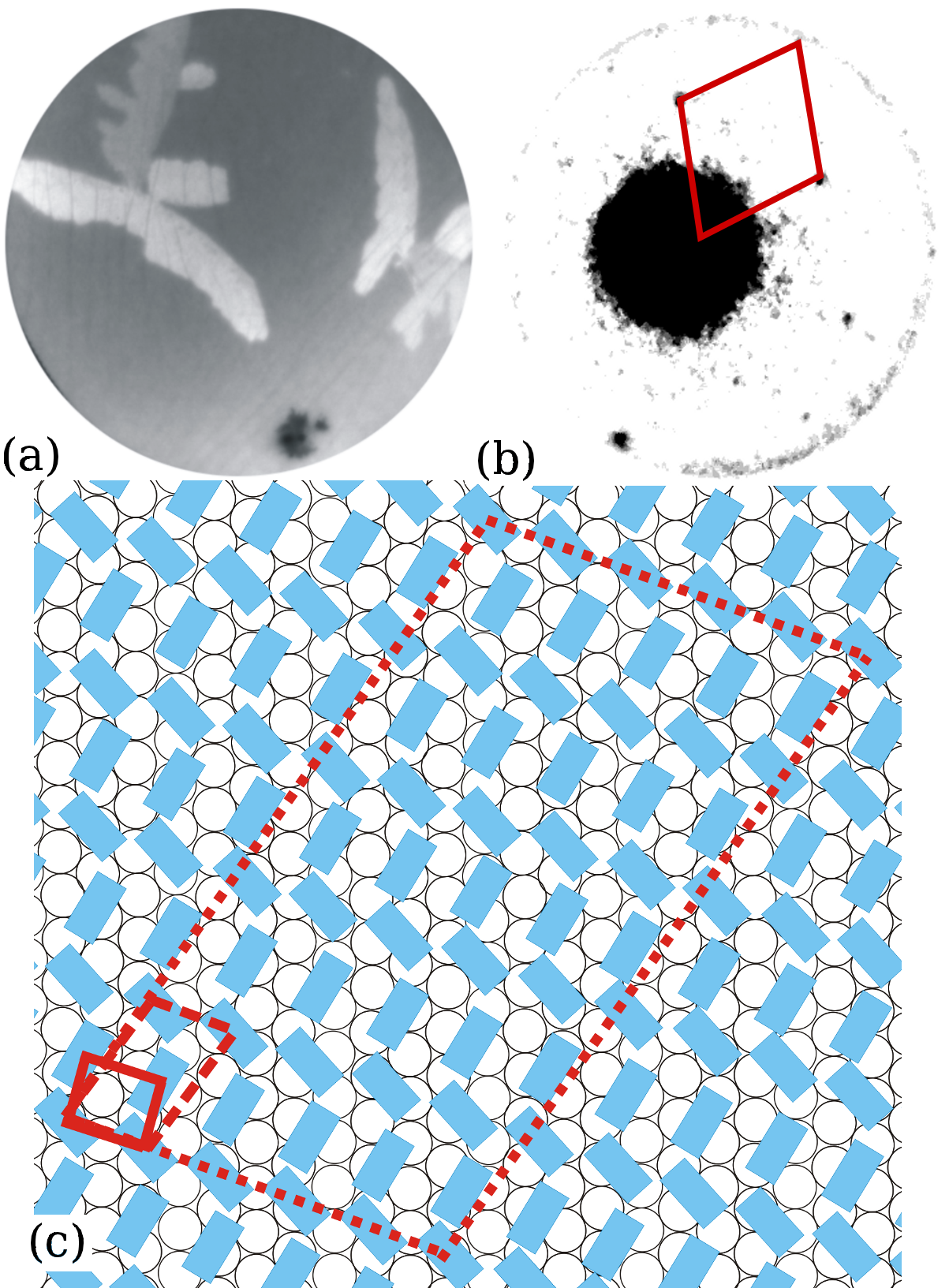}
   \caption{{\bf (a)} LEEM image of irregularly shaped structures of 6P grown
   on the Ir\{111\} surface. The Ir\{111\} surface appears dark and the ramified
   6P islands show different shades of grey. (FoV: 15\,$\mu$m, electron
   energy: 3.5\,eV, deposition temperature: 320\,K) {\bf (b)} $\mu$LEED pattern
   obtained from one of the islands at an electron energy
   of 19.4\,eV. The nearest neighbor cell is highlighted by red lines. {\bf (c)} The
   structural model proposed from the $\mu$LEED pattern shown in (b). The
   molecules are arranged in up-right standing orientation on Ir\{111\}.
   Nearest neighbor cell, unit cell and the 5$\times$5 superstructure
   are indicated by red lines (solid, dashed, and dotted respectively).
   }
   \label{fig3}
\end{figure}

Although the graphene flakes cover extended areas of the Ir\{111\} surface,
they still do not cover the entire surface. The remaining bare Ir\{111\} surface
areas are inspected after stopping the deposition of 6P molecules
(1.36\,6P/nm$^2$). LEEM
images show the presence of irregularly shaped 6P structures, as presented in
Fig.\,\ref{fig3}(a).
A $\mu$LEED measurement obtained from a branch of one of the irregularly
shaped structures is shown in Fig.\,\ref{fig3}(b). The $\mu$LEED pattern
reveals that 6P molecules form an ordered structure on the Ir\{111\} surface.
The dimensions of the nearest neighbor cell vectors are 5.0\,\AA{} by 5.0\,\AA{} at an
angle $\beta$ of 108$^\circ$. The size of this nearest neighbor cell implies that in
these irregularly shaped structures the long axis of the molecules is roughly
perpendicular to the surface. However, the cell vectors given above are the
nearest neighbor distances and not the real unit cell vectors. This is a
consequence of the molecular form
factors for the two differently rotated upright standing molecules being
nearly identical. \GH{The unit cell vectors are: 5.0\,\AA{} by 9.1\,\AA{} at an
angle $\beta$ of 105$^\circ$ and
$\Theta$=25$^\circ$
($\left(\begin{smallmatrix}3.8&1.6\\-0.4&1.6\end{smallmatrix}\right)$).
   Considering the above mentioned measurement precision and the
   fact that some of the molecules will shift slightly to reach a
   more favorable position, a 5$\times$5 superstructure with a matrix
   notation of
   $\left(\begin{smallmatrix}19&8\\-2&8\end{smallmatrix}\right)$ (a
      coincidence type II quasiepitaxial relationship~\cite{Hooks2001})
      describes the situation more accurately. This can be seen in
      fig.~\ref{fig3}(c) where some of the molecules would need to be
      shifted only slightly by fractions of an \AA{}ngstrom to reach a well
      coordinated site.} The 6P molecules are arranged in a similar
(up-right standing) fashion as in the (100) plane of the 6P bulk crystal.
Different 6P islands or arms of them can have different azimuthal
crystallographic orientations. This has been made visible in
Fig.~\ref{fig3}(a) by using a slightly off normal incident of the electron
beam. As a result different
crystallographic orientations show different intensities similar to a dark
field image. $\mu$LEED patterns recorded away from the irregular structures
consist only of Ir\{111\} spots and a dominant diffuse background. The latter is
attributed to an unordered 2D gas phase layer of 6P present on the surface
of the Ir.

It is well known that on clean metal surfaces para-n-phenyl oligomers prefer
a lying configuration~\cite{Winter2003,Mullegger2006,Hlawacek2004}.
However, small amounts of surfactants will lead to an upright standing
configuration of the
molecules~\cite{Hlawacek2004,Mullegger2006,Resel2005,Winter2006}. Therefore,
it is reasonable to assume that carbon residues of the graphene growth are
causing the appearance of these irregularly shaped structures on Ir\{111\}. 

\begin{figure}
   \centering
   \includegraphics[width=7.5cm]{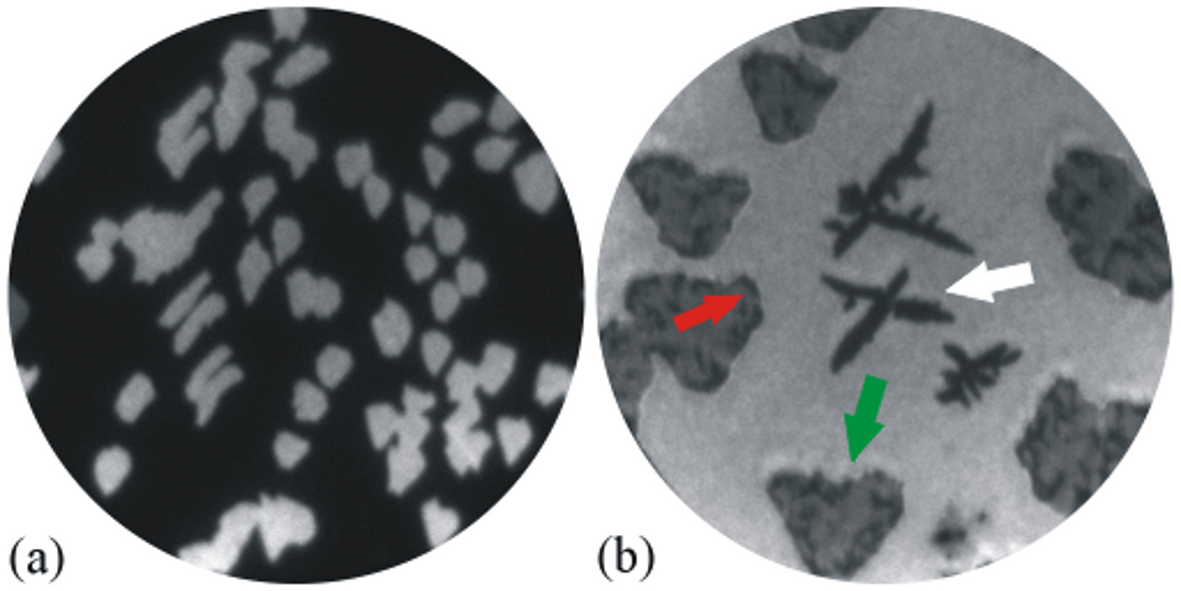}
   \caption{{\bf(a, t = 205\,s)} 100\,$\mu$m FoV PEEM image of Ir\{111\}
   covered with graphene flakes. The Ir\{111\} surface appears dark, since its
   work function (5.76\,eV) is higher than the photon energy (4.9\,eV).
   {\bf(b, t = 2149\,s)} 50\,$\mu$m FoV PEEM image acquired after 6P
   deposition at a temperature of 320\,K. The 6P structures, graphene
   flakes, 
   and ramified islands on Ir\{111\} are marked with red, green, and white
   arrows, respectively. Times indicated are measured with
   respect to the start of ethylene (C$_2$H$_4$) and 6P deposition,
   respectively.}
   \label{fig1b}
\end{figure}

PEEM relies on photo-emitted electrons and therefore depends on changes in
the work function of a sample to create image contrast. The clean Ir\{111\}
surface appears dark, since its work function (5.76 eV~\cite{Strayer1973}) is higher than the photon
energy (4.9\,eV) whereas the graphene (4.8\,eV-4.9\,eV~\cite{Starodub2011})
flakes appear bright (Fig.\,\ref{fig1b}(a)). However, after deposition of 6P
the Ir\{111\} surface appears brighter than graphene (Fig.\,\ref{fig1b}(b)).
The change in contrast is suggestive of a surface work function variation
caused by 6P adsorption and the formation of an interface dipol -- both on Ir\{111\} and graphene. The 6P needles grown
on graphene (indicated by a red arrow) appear darker than the 6P wetting layer
on the graphene (Fig.\,\ref{fig1b}(b). A white arrow is indicating the
irregularly shaped structures on the Ir\{111\} surface which gives a relatively
darker contrast. The 6P covered graphene flakes appear darker than Ir\{111\}
and have lighter shade of grey than the 6P needles. Therefore, the resulting
order in brightness (from low to high) of the materials roughly grouped by work function
is: Ir\{111\} and upright standing 6P islands on Ir(111) (both higher or
similar to the photon energy), 6P$(\mathrm{1\overline{11})}$-needles,
6P$(\mathrm{1\overline{11}})$ wetting layer on graphene, disordered 6P on
Ir\{111\}. The non-emitting 6P needles are therefore only visible because they
sit on a brighter background. This is similar to the contrast mechanism
observed for the case of 6P/Cu(110)\,2$\times$1-O~\cite{Fleming2009}.

\subsection{Measurements at 352\,K}

\begin{figure}
   \centering
   \includegraphics[width=8cm]{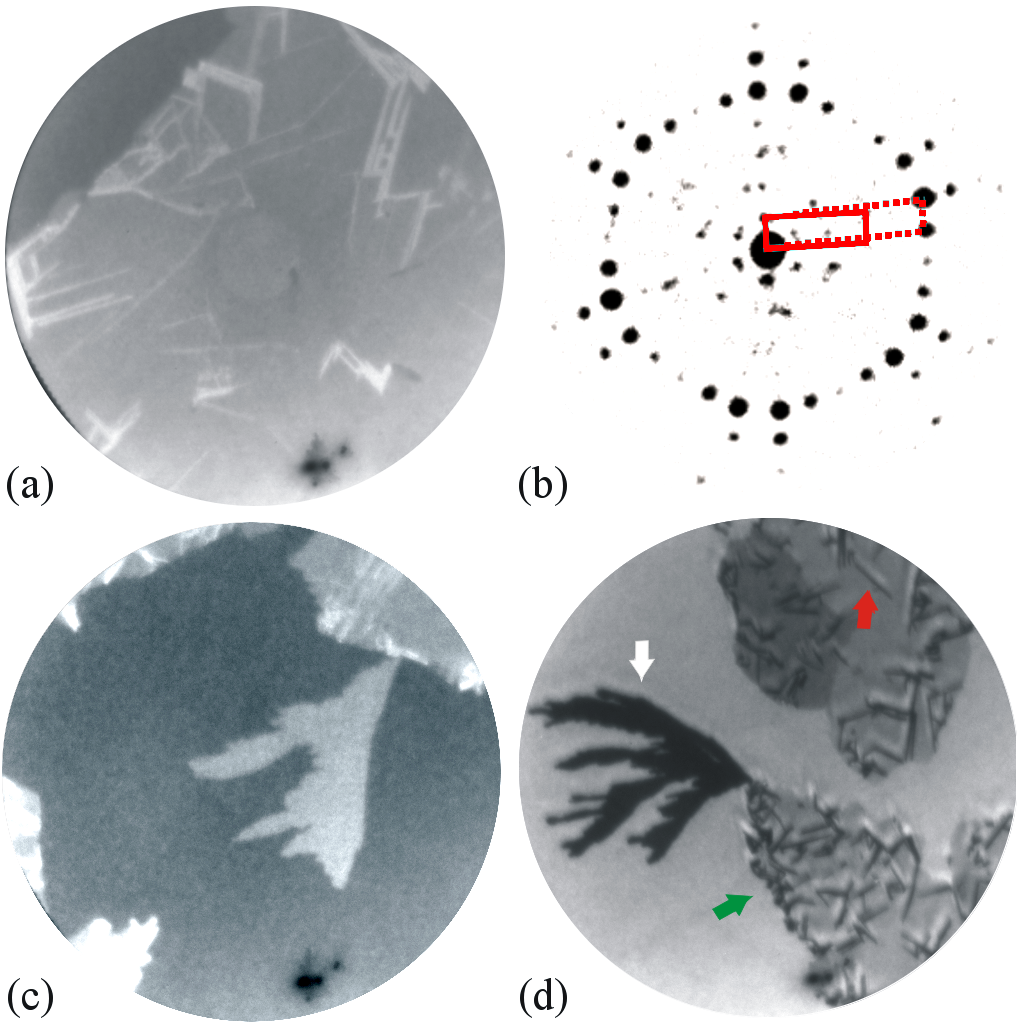}
   \caption{{\bf (a, t = 2130 s, 1.35\,6P/nm$^2$)} 10\,$\mu$m FoV LEEM image acquired at an
   electron energy of 2.7\,eV and 352\,K. A single graphene flake on the Ir\{111\}
   surface is imaged after deposition of 6P. The edge of the graphene flake is
   visible in the upper left part. The graphene flake is covered with 6P
   needles of different orientation. {\bf (b)} $\mu$LEED pattern measured
   from graphene covered by the wetting layer at an electron energy of
   19.3\,eV. {\bf(c)} 20\,$\mu$m FoV LEEM images
   acquired at an electron energy of 2.7\,eV and a temperature of 352\,K.
   The Ir\{111\} surface with an irregular shaped island and three graphene
   flakes covered with 6P is visible. The 6P island on the Ir\{111\} surface
   is connected to the graphene flake. {\bf (d)} 50\,$\mu$m FoV PEEM image 
   acquired after stopping the 6P deposition. 6P needles, graphene flakes, and 6P islands on Ir are
   present and marked by red, green, and white arrows, respectively (352\,K).
   }
   \label{fig5}
\end{figure}

Increasing the deposition temperature to 352\,K, leads to no principle
changes in the film formation process. After the initial two-step formation
of a wetting layer -- by nucleation of domains near the wrinkles -- the growth of
parallel needles sets in. Again the needles nucleate either
near the wrinkles, or from existing needles creating comb like 
structures (Fig.~\ref{fig5}(a)). As expected, higher deposition temperatures
and the resulting enhanced mobility of 6P leads to fewer, but longer
needles~\cite{Balzer2004}.

A typical $\mu$LEED pattern measured from the graphene surface covered by
the wetting layer is shown in Fig.~\ref{fig5}(b). The $\mu$LEED
pattern consists of the specular beam reflection surrounded by several rings of
LEED spots. This $\mu$LEED pattern is similar to the one obtained at
320\,K presented in Fig.~\ref{fig2}(a). The structure of
the wetting layer at this elevated temperature is identical to the
one that was already found for the growth at 320\,K. Due to the small signal, no
reliable structural information could be obtained from the 
needles. However, taking into account the similarities in the wetting layer
and the comparable morphology, one can conclude their structure is similar to
the structure at 320\,K presented in Fig.~\ref{fig2}(d).

Post-deposition (2130\,s, 1.35\,6P/nm$^2$) LEEM imaging of the Ir\{111\} surface reveals the presence of
branched 6P structures (Fig.\,\ref{fig5}(c)). All 6P structures on Iridium
nucleate at the edges of the graphene flakes. The increased mobility of 6P
on Ir\{111\} at this high temperature requires the stable graphene flakes for
nucleation. Once formed, they act as sinks for all 6P diffusing on the
Ir\{111\} surface. A similar structure of upright molecules as observed for
the other deposition temperatures is proposed. 

A PEEM image acquired after stopping the deposition of 6P is shown in
Fig.~\ref{fig5}(d). The 6P needles on the graphene flake (indicated with a red arrow)
appear darker than the 6P wetting layer in the same way as described above.
A white arrow marks the irregular and branched structures on the
Ir\{111\} surface. Again, they show a darker contrast than the surrounding
surface. The 6P wetting layer on the graphene flakes itself shows an
intermediate grey level.   

\begin{figure} 
   \centering 
   \includegraphics[width=7cm]{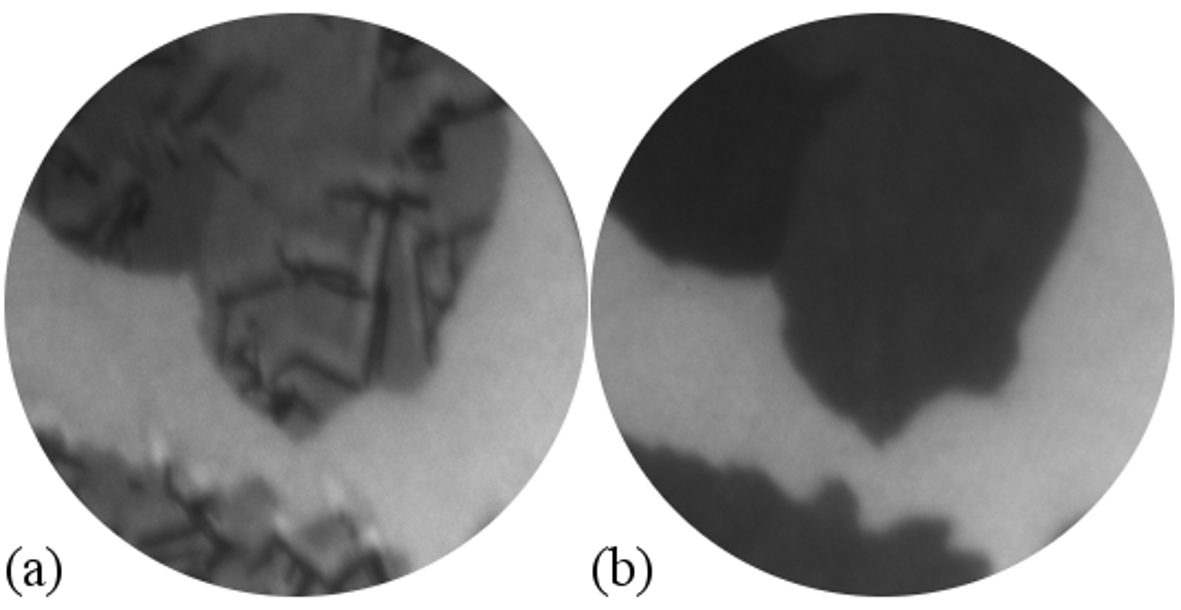}
   \caption{{\bf(a, T = 381\,K)} PEEM image acquired after stopping the
   6P deposition. The 6P needles, on two graphene flakes can be seen.
   {\bf(b, T = 401\,K)} The same two
   graphene flakes are cleared from all 6P needles. (FoV: 26\,$\mu$m)} 
   \label{fig6a} 
\end{figure}

Post deposition annealing of the film leads to a decay of the structures.
From deposition temperature to 381\,K 6P structures on graphene and Ir\{111\}
remain intact and immobile (Fig.~\ref{fig6a}(a)). With a further increase of temperature, first the
small and later also the bigger needles start to decay until at 400\,K all
structures on the flakes have disappeared (Fig.~\ref{fig6a}(b)). The excess molecules can diffuse off the graphene
flake into the 2D gas phase on the supporting Ir\{111\} substrate. A further increase of
temperature results in a shrinking of the -- so far unchanged -- irregularly shaped
structures on the Ir\{111\} surface. They eventually disappear all at 416\,K. When comparing these results to
desorption data obtained on other substrates~\cite{Winter2004,Mullegger2006},
uncertainties of the temperature measurements in the LEEM sample holder as well as the low heating rate of
only 6\,K/min have to be taken into account.
The sequence in which 6P desorbs from the different substrates is further
evidence underlining the weak interaction of 6P with graphene.

\subsection{Measurements at 405\,K}

\begin{figure}
   \includegraphics[width=7.6cm]{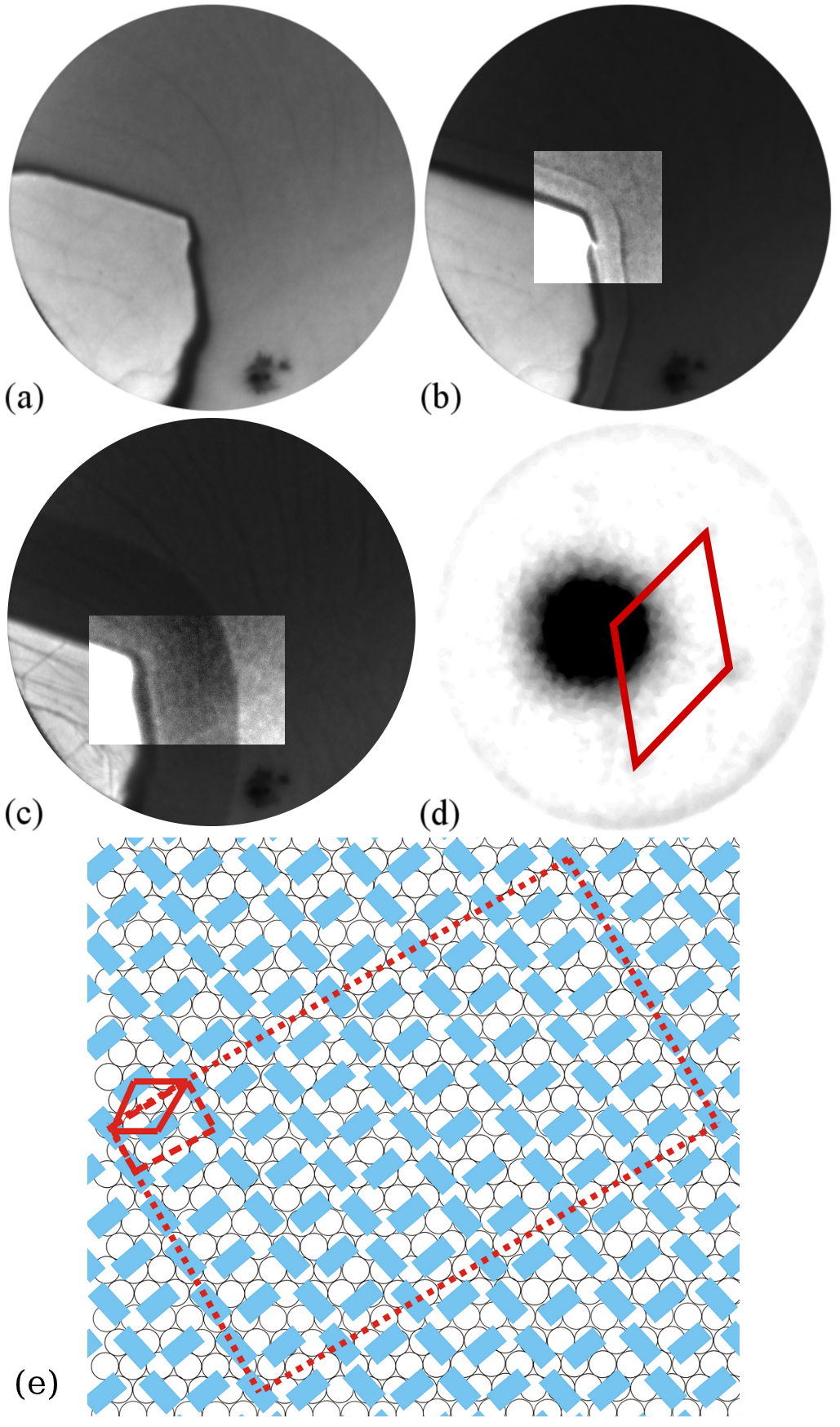}
   \caption{6\,$\mu$m FoV LEEM images acquired at an electron energy of
   2.7\,eV and temperature of 405 K.  {\bf (a, t\,=\,0\,s)} A graphene flake
   residing on the Ir\{111\} surface prior to exposure to 6P. Wrinkles and the
   contours of Ir\{111\} surface steps are visible on the single layer
   graphene flake. {\bf (b, t\,=\,831\,s, 0.53\,6P/nm$^2$)} The nucleation of a 6P film takes
   place on the edges of the graphene flake, as is witnessed by the
   appearance of a band of different intensity at the edge of the graphene
   (see contrast enhanced center).
   The reflected intensity from the Ir\{111\} surface decreases. Three
   different levels of contrast are found. From left to right: graphene,
   ordered 6P layer and 6P lattice gas on Ir\{111\}. 
   {\bf (c, t\,=\,1391\,s, 0.88\,6P/nm$^2$)} The ordered 6P film extending from the graphene
   flake has grown further (see contrast enhanced center). {\bf (d)} A
   $\mu$LEED pattern is measured at electron energy of 25.6\,eV using a
   1.4\,$\mu$m field-limiting aperture from the Ir\{111\} surface area completely
   covered with the ordered 6P layer. The nearest neighbor cell is
   highlighted by red lines. {\bf (e)} The structural model proposed from
   the $\mu$LEED pattern shown in panel (d). The molecules are arranged in
   an up-right standing orientation on Ir\{111\}. The unit cell (dashed lines)
   and the nearest neighbor cell (solid line) are shown. The 6$\times$6
   superstructure is indicated by a dotted line. Times indicated 
   are measured with respect to the start of 6P deposition.}
   \label{fig7}
\end{figure}

Fig.~\ref{fig7} is a sequence of images recorded during 6P deposition at 405\,K. 
Fig.~\ref{fig7}(a) shows the initial situation. The reflected intensity from the
Ir\{111\} surface decreases with deposition time, indicating the presence of a
diluted phase of
6P on the surface. However, at this elevated temperature neither the formation
of a wetting layer, nor the nucleation of any other 6P structure is observed
on graphene. We believe, that the already large diffusion length of 6P at
lower
temperatures (i.e. as low as 240\,K~\cite
{Hlawacek2011}), will be of the order of the radius of the graphene
flakes (roughly 2\,$\mu$m) at 405\,K. As a result,
the 6P molecules diffuse from the flakes onto the Ir\{111\} surface, where
6P domains nucleate at the edges of the graphene flake. This process begins
after 230\,s (0.14\,6P/nm$^2$) of deposition (Figs.~\ref{fig7}(b) and
(c)). The contrast enhanced centers of fig.~\ref{fig7}(b,c) allow to distinguish
between the graphene flake (left and brightest), ordered 6P film and 6P gas phase (upper right). However,
the borders between the different areas (in particular in fig.~\ref{fig7}(b)
are affected by a LEEM image artefact related to abrupt changes in
morphology and workfunction~\cite{Schertz2011}. 

A $\mu$LEED pattern obtained from the dark band next to graphene flake
in fig.\ref{fig7}(c) on the 6P covered Ir\{111\} surface is shown in
Fig.\,\ref{fig7}(d).  Only very diffuse spots can be found, superimposed on a
homogeneous, diffuse background. The crystalline quality of this film is not
very high. The nearest neighbor cell highlighted in Fig.~\ref{fig7}(d) has
a size of 5.0\,\AA{} by 5.0\,\AA{} with an angle $\beta$ of 120$^\circ$.
The obvious way to accommodate the 6P molecules into such a small space is
in an upright standing way where the long molecular axis is roughly
perpendicular to the substrate. Using the unit cell of the bulk (100) plane
(8.091\,\AA{} by 5.568\,\AA{} and $\beta$=90$^\circ$~\cite{bakandfra93}) as a
starting point we can deduct the unit cell of 6P on Ir\{111\} to be 8.7\,\AA{}
by 5\,\AA{} and $\beta$ and $\Theta$=90$^\circ$ (dashed line in Fig.~\ref{fig7}(e)). Compared
to the bulk structure this unit cell is compressed along the short axis.
\GH{The resulting
matrix notation of the overlayer with respect to the underlying Ir\{111\} is
given by the following quasiepitaxial coincidence type II
relationship~\cite{Hooks2001} $\left(\begin{smallmatrix}1.9& 3.7\\1.9&
   0\end{smallmatrix}\right)$. Using the same arguments as for the previous
   structures a 6$\times$6 superstructure describes the situation more
   precisely and results in the following matrix notation$\left(\begin{smallmatrix}11&22\\
   11&0\end{smallmatrix}\right)$  which is depicted in
   fig.~\ref{fig7}(e).} The distortion of the 6P unit cell is geometrically justified
as the molecular rows will have the substrate dictated 120$^\circ$ angle. $\mu$LEED patterns obtained far away from the flakes
show only the unordered 2D gas phase of 6P. Different to the well
investigated~\cite{Resel2006a,Hlawacek2005,Berkebile}, but non-metallic
system -- 6P on TiO$_2$ -- we see no evidence for
an additional ordered layer of flat lying molecules~\cite{Sun2010}.

In general, increased substrate temperatures have been identified as one
of the reasons for the growth of up-right standing 6P
molecules~\cite{AFR+99,Haber2006a,Mullegger2007}.
In the
same way, the elevated surface temperature of Ir\{111\} favors the growth
of up-right standing 6P thin films. No other
structures -- neither on Ir\{111\} nor on graphene -- were found for this
deposition temperature.

\section{Summary and conclusions}

The deposition of 6P molecules and growth of 6P structures on graphene has
been studied at different temperatures. For sample temperatures during
deposition up to 352\,K, 
wrinkles in the graphene act as preferential nucleation sites for both, a
$(\mathrm{1\overline{11}})$ wetting layer and 6P needles with the same
crystallographic orientation. The 6P needles form after the
completion of the wetting layer. This is usually identified as the
Stranski-Krastanov growth mode, often observed for 6P films formed from flat
lying molecules~\cite{Teichert2006}.

Defects of the Ir\{111\} substrate -- a result of carbon residues after the
formation of the graphene flakes -- are nucleation sites for
the growth of ramified structures consisting of upright standing 6P
molecules. However, with increasing sample temperature (compare
fig.~\ref{fig1}(d) at 320\,K to fig.~\ref{fig5}(a) at 352\,K), less but longer 6P
needles are formed on graphene. In addition 6P nucleation on the Ir\{111\}
surface gets increasingly difficult and ramified islands of upright 6P are
exclusively nucleated at the rim of graphene flakes. Further increase of the
deposition temperature to 405\,K results in a considerable change of growth behavior.
Neither a wetting layer, nor any three dimensional needles are observed on
graphene. A 6P(100) layer does, however, nucleate at the edges of the graphene
flakes. It grows on the Ir\{111\} surface in a step flow-like fashion. This
layer built from upright standing molecules shows poor crystallinity. 

Our study illustrates that at all temperatures investigated, the growth behavior of 6P on
graphene and Ir\{111\} is governed by defects. Up to 352\,K, graphene wrinkles dictate the nucleation and growth
behavior of the 6P wetting layer, and needles. At 
405\,K, the edges of the graphene flakes are the sites where 6P domains
develop on Ir\{111\}. 

\section*{Acknowledgement}

This work was financially supported by Austrian Science Fund (FWF) project
N9707-N20, STW, and FOM project 04PR2318.

\bibliographystyle{model1-num-names}
\bibliography{SurfSci}

\end{document}